\documentclass[12pt]{article}
\usepackage{graphicx}

\newcommand\pubdate{\today}

\def\lbnl{Nuclear Science Division\\
Lawrence Berkeley National Laboratory, Berkeley, CA 94720 USA and \\
\vskip .1 in
Physics Department, University of California \\
Berkeley, CA 94720 USA}
\def\support{\footnote{See http://icecube.wisc.edu/collaboration/authors/current for full author list and acknowledgements.}}

\def\Title#1{\begin{center} {\Large #1 } \end{center}}
\def\Author#1{\begin{center}{ \sc #1} \end{center}}
\def\Address#1{\begin{center}{ \it #1} \end{center}}

\newcommand\pubblock{\rightline{\begin{tabular}{l} 
         \pubdate  \end{tabular}}}
\newenvironment{Abstract}{\begin{quotation}  }{\end{quotation}}
\newenvironment{Presented}{\begin{quotation} \begin{center} 
             PRESENTED AT\end{center}\bigskip 
      \begin{center}\begin{large}}{\end{large}\end{center} \end{quotation}}
\def\Acknowledgements{\bigskip  \bigskip \begin{center} \begin{large}
             \bf ACKNOWLEDGEMENTS \end{large}\end{center}}

\def\beq{\begin{equation}}
\def\eeq#1{\label{#1}\end{equation}}
\def\eeqn{\end{equation}}

\def\beqa{\begin{eqnarray}}
\def\eeqa#1{\label{#1}\end{eqnarray}}
\def\eeqan{\end{eqnarray}}

\let\bar=\overbar

\def\Dslash{\not{\hbox{\kern-4pt $D$}}}
\def\dslash{\not{\hbox{\kern-2pt $\del$}}}

\def\msb{{\bar{\ssstyle M \kern -1pt S}}}

\begin{document}
\begin{titlepage}
\pubblock

\vfill
\Title{Astrophysical Neutrinos with IceCube}
\vfill
\Author{ Spencer R. Klein for the IceCube Collaboration\support}
\Address{\lbnl}
\vfill
\begin{Abstract}
The 1 km$^3$ IcCube neutrino observatory was built to find high-energy neutrinos that  are associated with the sources of ultra-high energy cosmic rays.  Its 5,160 optical sensors detect Cherenkov light from the charged particles produced when neutrinos interact in the ice.  

In this talk, I will describe the techniques that IceCube has used to search for astrophysical neutrinos.  An emphasis will be given to diffuse neutrinos (not associated with a specific source), including analyses of contained events and energetic through-going neutrinos.  At the end, I will discuss multi-messenger astronomy, and present an intriguing correlation between a high-energy IceCube neutrino and a blazar in  outburst. 


\end{Abstract}
\vfill
\begin{Presented}
Thirteenth Conference on the Intersections of Particle and Nuclear Physics \\
May 29-June 3, 2018
\end{Presented}
\vfill
\end{titlepage}
\def\thefootnote{\fnsymbol{footnote}}
\setcounter{footnote}{0}

\section{Introduction}

The origin of ultra-high energy cosmic-rays is an unsolved mystery.  Somewhere, astrophysical particle accelerators accelerate protons or heavier ions to energies above $10^{20}$ eV.  Unfortunately, nuclear cosmic-rays are bent by interstellar magnetic fields, so their arrival directions on Earth do not point back to their sources.  Despite more than 60 years of studies of ultra-high energy cosmic rays, we have not yet found definitive evidence of any specific source or source classes.    One way to find these accelerators is to search for them using a different type of particle: the neutrino.  Neutrinos are electrically neutral, so travel in straight lines, and they have  small enough interaction cross-sections to escape from even dense sources.    They can be produced when nuclei undergoing acceleration interact with either gas or photons in or near the accelerator.   The number of neutrinos depends on the density of the gas or photons. 

Because they interact so weakly, a large detector is needed to observe astrophysical neutrinos.  Two types of calculations have been used to estimate the neutrino flux, and, from that the required detector size.    One used the measured cosmic-ray flux and estimates of the beam-gas or beam-photon density.   The maximum neutrino flux occurs when the source is just dense enough to absorb all of the energy from the proton beam; this is known as the Waxman-Bahcall limit \cite{Bahcall:1999yr}.   The other used the measured gamma-ray flux, assuming that the gamma-rays come from $\pi^0\rightarrow\gamma\gamma$.  Both calculations found that a 1 km$^3$ detector should be large enough to find astrophysical neutrino sources.  These results drove the size of the IceCube neutrino detector.

IceCube consists of 1 km$^3$ of Antarctic ice at the South Pole, instrumented with 5,160 digital optical modules (DOMs) \cite{Aartsen:2016nxy}.   The DOMs observe the Cherenkov radiation emitted by the charged particles that are produced when neutrinos interact in the ice \cite{Halzen:2010yj}.  They are deployed on 86 vertical strings, each holding 60 DOMs.  78 of the strings are distributed on a 125 m triangular grid, covering about 1 km$^2$, with the DOMs spaced every 17 meters between 1450 and 2450 m below the surface.  The remaining 8 strings, called ``Deep Core" are deployed near the center of the array.  They have smaller string-to-string and DOM-to-DOM spacings, giving Deep Core a lower energy threshold than the rest of the detector.  IceCube was constructed between 2005 and 2010 by an international collaboration.

Each DOM consists of a 25.4 cm photomultiplier tube in a glass pressure vessel, along with data acquisition, calibration and communications systems \cite{Abbasi:2008aa}.  The DOMs operate autonomously, receiving power and control signals and sending packetized digital data to the surface.  A calibration system exchanges pulses with the surface, maintaining the DOM-to-DOM timing calibrations within 3 nsec.  Thirteen on-board LEDs are used for PMT and inter-DOM calibrations, including to measure the optical properties of the Antarctic ice \cite{Aartsen:2013rt}. 

\section{Atmospheric muons and neutrino backgrounds}

Astrophysical neutrino searches must contend with two types of backgrounds.  The first are downward-going cosmic-ray muons, which are produced in cosmic-ray air showers.  These are far more numerous than neutrinos, with IceCube triggering at about 2800 Hz, mostly from these muons.   Two ways to avoid this background are to select upward-going tracks, since the Earth acts as a muon shield, or to select interactions that originate within the detector, with no sign of an incident track.  Because atmospheric neutrinos are produced in cosmic-ray air showers, they are likely to be accompanied by air shower particles and cosmic-ray muons; these additional particles may be used to veto atmospheric neutrinos. 

\begin{figure}[t]
\includegraphics[height=2.2in]{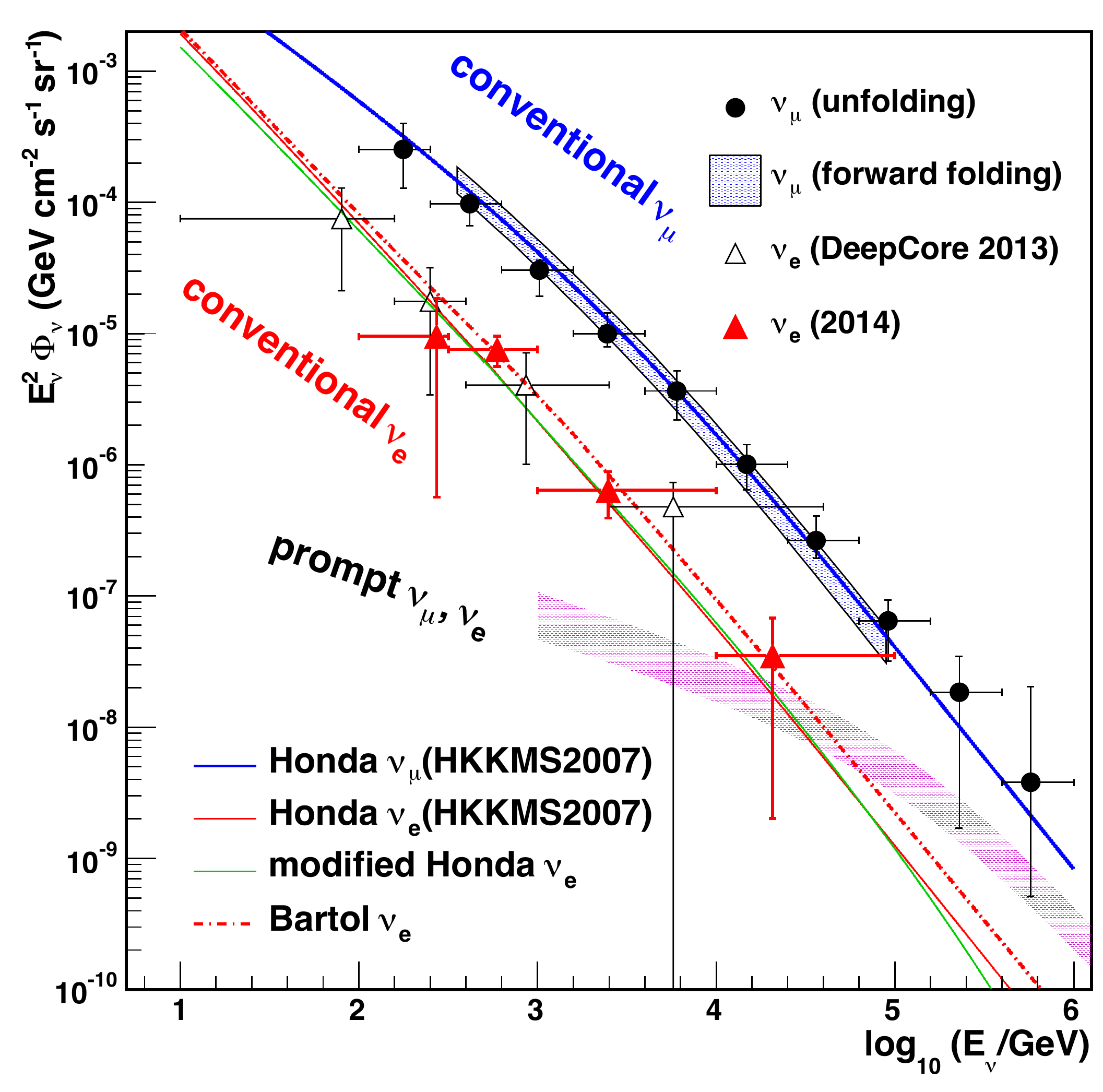}
\includegraphics[height=2.2in]{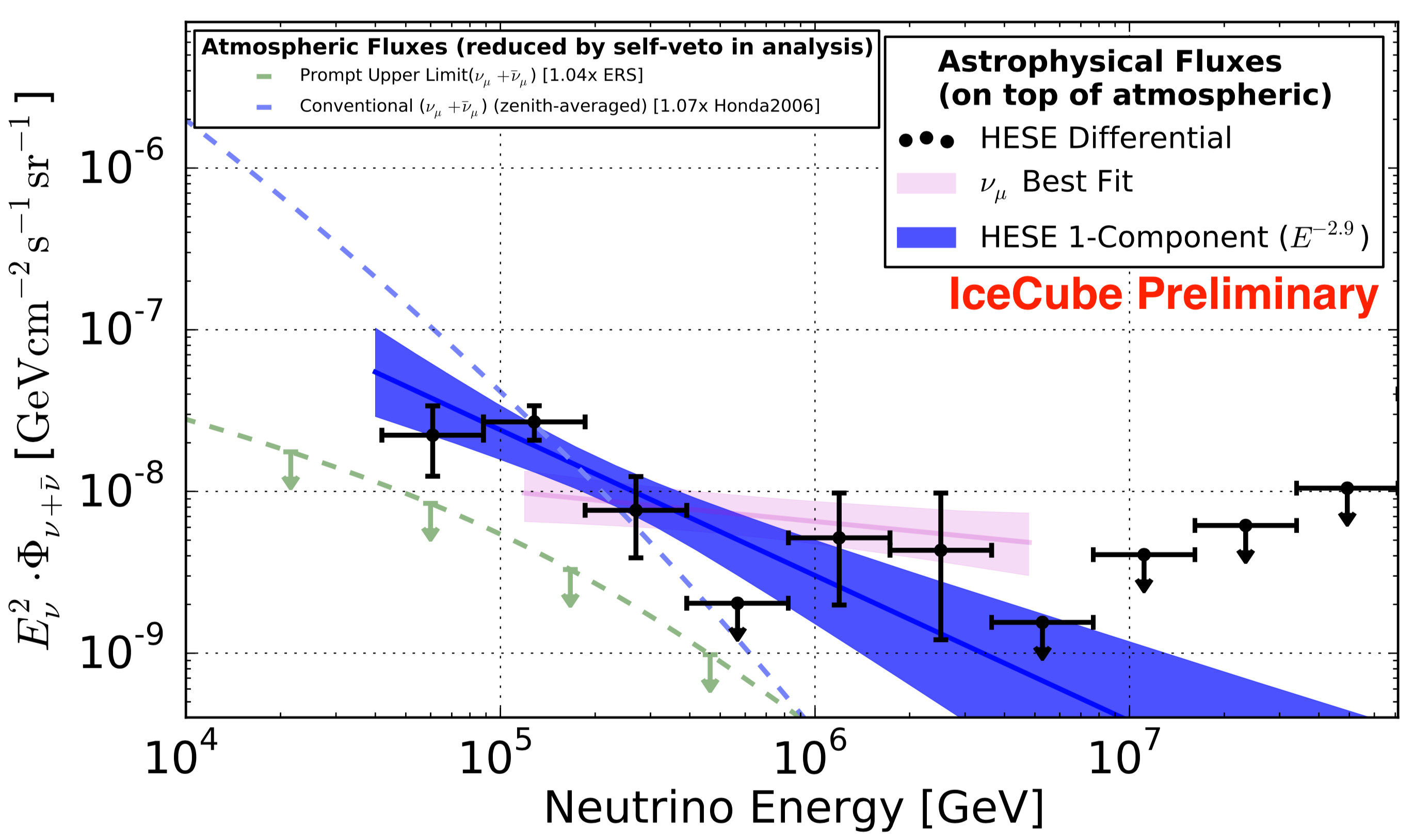}
\caption{(Left) Atmospheric $\nu_e$ and $\nu_\mu$ flux measurements, compared with the calculations used by IceCube.  The red band shows the prompt flux, which has not yet been observed. From Ref. \cite{Aartsen:2015xup}.
(Right) The measured differential astrophysical flux measured using contained events (points) and a fit to that data (blue/purple line and band), compared with the best fit obtained from through-going $\nu_\mu$ (pink line and band).  From Ref. \cite{Aartsen:2017mau}.
}
\label{fig:atmos}
\end{figure}

The second background is from atmospheric neutrinos, neutrinos produced in cosmic-ray air showers.    Conventional atmospheric neutrinos come from the decay of pions and kaons.  These neutrinos are mostly $\nu_\mu$, with slightly more $\nu$ than $\overline\nu$.  Because IceCube cannot generally differentiate between $\nu$ and $\overline\nu$, we will not further distinguish between them here.   The conventional atmospheric neutrino energy spectrum depends on the cosmic-ray air shower spectrum.  Pions and kaons are relatively long-lived, so they may interact in the atmosphere before they can decay.   This softens the neutrino energy spectrum, leading to a conventional neutrino energy spectrum that roughly goes as $dN_\nu/dE_\nu \propto E_\nu^{-3.7}$ for, roughly, $E_\nu< 100$ TeV, softening to $dN_\nu/dE_\nu \propto E_\nu^{-4.0}$ at higher energies.  This competition also affects the angular distribution; conventional atmospheric neutrinos are concentrated around the horizon.  IceCube uses calculations of particle production in the atmosphere to model neutrino production \cite{Honda:2006qj}.  As Fig. \ref{fig:atmos} shows, the  calculations are in good agreement with the $\nu_\mu$ \cite{Aartsen:2017nbu} and $\nu_e$ \cite{Aartsen:2015xup,Aartsen:2012uu} data.  

Prompt atmospheric neutrinos come from the decay of charm and bottom quarks.  Because they are short-lived, they are unlikely to interact, so they follow the cosmic-ray air shower energy spectrum, and they are nearly isotropic.  Prompt neutrinos have yet to be observed, but IceCube has set flux limits that are not too much higher than the theoretical predictions. 

\section{Finding Astrophysical Neutrinos}

Many approaches have been used to search for a small flux of astrophysical neutrinos above these large backgrounds.  One approach is to 
search for point sources, which produce a local concentration which sticks up above the smooth backgrounds.  Another is to search for particularly energetic neutrinos, since astrophysical neutrinos are expected to have a spectrum that roughly goes as $dN_\nu/dE_\nu \propto E_\nu^{-2.0}$, harder than the atmospheric neutrinos.  A third approach is to search for downward-going neutrinos that are unaccompanied by cosmic-ray air showers and atmospheric muons.  Finally, IceCube is also searching for $\nu_\tau$, which are very rare in air showers, coming only from $D_s^+$ and $B$-meson decays.  In contrast, they are expected to be 1/3 of the astrophysical flux, since neutrino oscillations in-transit convert $\nu_\mu$ and $\nu_e$ into $\nu_\tau$.  IceCube has yet to see a clear $\nu_\tau$ signal.

The first hint of an astrophysical signal came from two neutrinos, Bert and Ernie, that were found in a search for extremely-high energy neutrinos \cite{Aartsen:2013bka}.  Each were well-contained cascades, with an energy around 1 PeV - essentially golden events.   The predicted atmospheric background was $0.082\pm 0.004$ (stat.) $^{+0.06}_{-0.04}$ events.   

These events prompted a search for more events that originated within the detector.  This ``High-Energy Starting Event" (HESE) analysis divided the detector into an outer veto region, covering the top 10 DOMs in most strings, all of the outer strings, and the bottom DOMs in most strings, and a signal region.  A dusty layer in the middle of the detector and a surrounded buffer were also included in the veto region, to eliminate muon tracks that entered undetected.    It selected events that deposited more than 6,000 observed photons (photoelectrons) in the detector, but where the first significant deposition was in the signal region.  The two year HESE search found strong evidence for astrophysical neutrinos, while the three-year search crossed the $5\sigma$ discovery threshold.  Here, I discuss a newer search which found 82 events in 2078 live days (over 6 years) of data \cite{Aartsen:2017mau}.  The expected background from downward-going muons was $25\pm 7$ events, determined by adding a second, inner veto layer, and comparing the pass rates in the two layers.  The estimated conventional atmospheric neutrino background was $16^{+11}_{-4}$ events, including prompt neutrinos, which were constrained by a previous IceCube $\nu_\mu$ study  \cite{Aartsen:2013eka}.

An additional cut to remove most of the muon background required that events have more than 60 TeV of energy deposited in the detector.  The astrophysical component was then fit to a power law $dN_\nu/dE_\nu \propto E_\nu^{-\gamma}$, and the best fit value $\gamma= 2.92\pm 0.3$ was found. This index is somewhat softer than expected; most models based on Fermi acceleration predict $\gamma \approx 2$.  The arrival distribution of the events was consistent with isotropy, as expected.

An independent search for astrophysical neutrinos used energetic upward-going, through-going muons.   Through-going muons offer very good angular resolution, but poor energy resolution, because of the uncertainty about how far outside the detector the neutrino interacted.   A fit to the measured muon energy spectrum found a clear $5.6\sigma$ excess over atmospheric expectations \cite{Aartsen:2016xlq} with a best-fit spectral index $\gamma = 2.2\pm 0.1$.  The index is consistent with Fermi acceleration, but in significant tension with the HESE sample.  However, as Fig. \ref{fig:atmos} shows, the through-going muon analysis samples more energetic neutrinos than the HESE analysis.   In the energy region where the two samples overlap, the flux estimates agree reasonably well.   One possible  explanation is that the energy spectrum is not a single power law.  However, fits to the HESE sample do not show a preference for a double power law.  Another possible explanation is that the track and cascade energy spectral indices are different; this would likely point to an energy-dependent acceleration mechanism,  a non-standard oscillation scenario, or a non-standard acceleration scenario.

\begin{figure}[htb]
\centering
\includegraphics[height=2.2in]{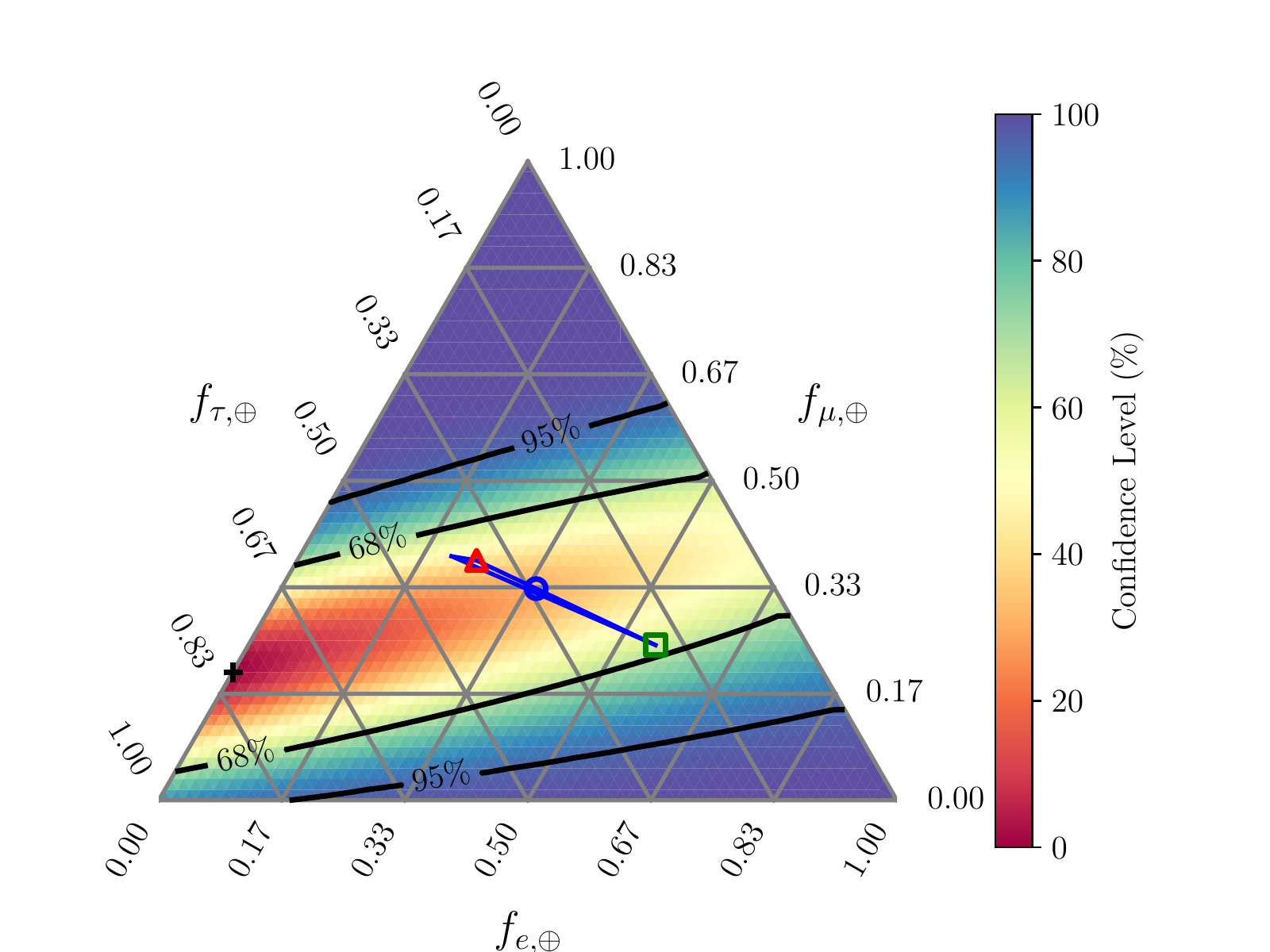}
\includegraphics[height=2.2in]{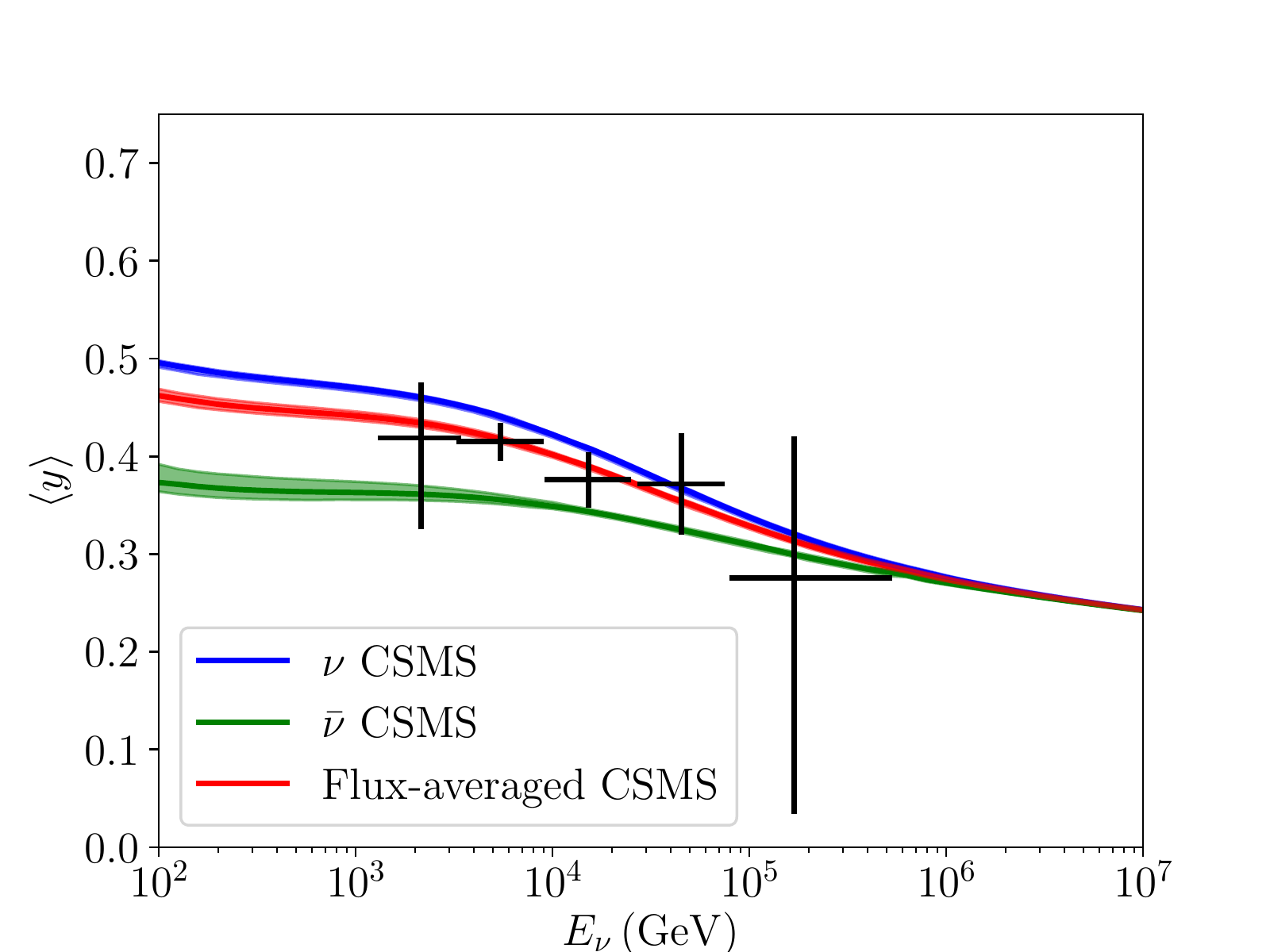}
\caption{(Left) The neutrino flavor triangle, showing the allowed regions based on the comparison of starting tracks and cascades. (Right) The measured mean inelasticity as a function of neutrino energy, compared with the standard-model expectations. 
Both from Ref. \cite{Aartsen:2018vez}. 
 } 
\label{fig:flavor}
\end{figure}

To study these possibilities, we performed two tests on a separate sample of contained events \cite{Aartsen:2018vez}. It consisted of 2650 starting tracks and 965 cascade events, selected using slightly different criteria, to extend the analysis to lower energies, while still removing atmospheric muon background.  The starting tracks were subjected to a new energy reconstruction method which used machine learning to separately reconstruct the cascade and track energies, and, from them, determine the visible inelasticity.    The inelasticity is discussed below, but here we present two astrophysical neutrino results.  First, we perform a fit to the conventional, prompt and astrophysical neutrino fluxes, but we allow the astrophysical fluxes to vary separately in the cascade and track samples.  This fit finds $\gamma=2.43^{+0.28}_{-0.30}$ for the track fit and $\gamma= 2.62\pm 0.08$ for the cascades, compared to $\gamma=2.62 \pm 0.07$ for the combined sample.  The cascade and combined-sample spectral indices are consistent with previous measurement, while the track spectrum is in between the HESE and through-going muon results, but with large enough error bars to encompass both.  A second fit to the sample allowed the astrophysical flavor ratio to vary from the standard $\nu_e:\nu_\mu:\nu_\tau = 1:1:1$.  Figure \ref{fig:flavor} (left) shows the result of this fit, showing the relative likelihood of flavor ratios.   100\% $\nu_\mu$ and 100\% $\nu_e$ are ruled out at greater than $5\sigma$, but the analysis cannot differentiate between different standard acceleration models. This study uses tracks and cascades of comparable energies, so is less dependent on the neutrino energy spectrum than the previous global fit \cite{Aartsen:2015knd}.

\section{Point source searches}

IceCube has made many searches for neutrinos from different point sources, and from different classes of objects.  So far, the only statistically significant positive result is from the blazar TXS0506 +56 \cite{IceCube:2018dnn}.  On Sept. 22, 2017, IceCube observed a neutrino which was energetic enough to be of likely astrophysical origin.  So, it issued a rapid alert response, which led several observatories to perform targeted observations in that direction.  Data from the Fermi telescope showed that the position was consistent with a known blazar which was emitting gamma-rays with an energy above 1 GeV.  The blazar was in an active state when the neutrino was observed, with higher than average gamma-ray emission.  Observations from the MAGIC telescope showed that the source was also emitting photons with energies above 100 GeV.    A subsequent search in archival IceCube data showed that the source has emitted a burst of neutrinos  during the period Sept. 2014 to March, 2015 \cite{IceCube:2018cha}.     Although confirmation is needed, it appears that we have finally located at least one astrophysical particle accelerator.  

\section{Particle and Nuclear Physics with IceCube}

IceCube can use the energetic neutrinos that it observes to study neutrino interactions, at energies far above the reach of particle accelerators.  One analysis used the zenith angle distributions of neutrinos observed in IceCube to study neutrino absorption in the Earth, and, from that, measure the neutrino-nucleon cross-section \cite{Aartsen:2017kpd}.  It used a two-dimensional fit to the zenith angle and muon energy distribution, where the cross-section was a free parameter in the fit.  Neutral-current interactions were included by treating absorption as a two-dimensional problem: neutrino energy entering the Earth, and neutrino energy observed in IceCube.  Near-horizontal neutrinos provided a nearly absorption-free baseline.    The fit assumed that the charged-current and neutral-current interactions were a single multiple, $R$ of the standard model cross-sections \cite{CooperSarkar:2011pa}, and found $R=1.30^{+0.21}_{-0.19}$ (stat.)$^{+0.39}_{-0.43}$ (syst.) for energies from 6.3 to 980 TeV.

A second analysis used the starting-track study mentioned above to measure the inelasticity distribution of neutrinos with energies from 1 TeV up to above 100 TeV \cite{Aartsen:2018vez}.  Figure \ref{fig:flavor} (right) shows the mean inelasticity, $\langle y \rangle$ as a function of energy.  The mean inelasticities are in good agreement with the standard model predictions \cite{CooperSarkar:2011pa}.

\section{Conclusions}

IceCube has measured a strong diffuse neutrino flux.  In the energy region where they overlap, two independent methods show some tension in the spectral index, but give similar flux measurements.   We have observed a coincidence between an energetic, likely-astrophysical neutrino from the direction of the blazar TXS0506+56, emitted during a time when the blazar was in outburst.   Using archival data, we then found one period, from Sept. 2014 to March, 2015, when the source was emitting a significant flux of neutrinos.   Together, the two observations point to this blazar as an astrophysical particle accelerator.

IceCube has used its sample of high-energy neutrinos to study neutrino interactions at energies far above those accessible at accelerators.   Two analyses have measured the neutrino-nucleon cross-section and the inelasticity distribution for charged-current interactions and found them in good agreement with the standard model.

Looking ahead, we expect to collect more data and extend these studies to higher energies with a future, Gen-2 upgrade \cite{Ackermann:2017pja}.

\Acknowledgements
This work was supported in part by U.S. National Science Foundation under grants PHY-1307472 and the U.S. Department of Energy under contract number DE-AC02-05-CH11231.

\end{document}